\documentclass{article}
\usepackage{spconf,amsmath,graphicx}

\usepackage{booktabs}
\usepackage{multirow}
\usepackage{url}

\title{Enhancement of Spatial Clustering-Based Time-Frequency Masks using LSTM Neural Networks}
\name{{Felix Grezes$^1$, Zhaoheng Ni$^1$, Viet Anh Trinh$^1$, Michael Mandel$^{12}$}}
\address{
  $^1$The Graduate Center, City University of New York\\
  $^2$Brooklyn College, City University of New York \\
\{fgrezes,zni,vtrinh\}@gradcenter.cuny.edu, mim@sci.brooklyn.cuny.edu}
\begin{document}
\ninept
\maketitle
\begin{abstract}




Recent works have shown that Deep Recurrent Neural Networks using the LSTM architecture can achieve strong single-channel speech enhancement by estimating time-frequency masks. However, these models do not naturally generalize to multi-channel inputs from varying microphone configurations. In contrast, spatial clustering techniques can achieve such generalization but lack a strong signal model.
Our work proposes a combination of the two approaches. By using LSTMs to enhance spatial clustering based time-frequency masks, we achieve both the signal modeling performance of multiple single-channel LSTM-DNN speech enhancers and the signal separation performance and generality of multi-channel spatial clustering.
We compare our proposed system to several baselines on the CHiME-3 dataset. We evaluate the quality of the audio from each system using SDR from the BSS\_eval toolkit and PESQ. We evaluate the intelligibility of the output of each system using word error rate from a Kaldi automatic speech recognizer.
\end{abstract}
\section{Introduction}
\label{sec:intro}
With speech recognition techniques approaching human performance on noise-free audio with a close-talking microphone \cite{achieving-human-parity-conversational-speech-recognition-2}, recent research has focused on the more difficult task of speech recognition in far-field, noisy environments. This task requires robust speech enhancement capabilities.

One approach to speech enhancement is spatial clustering, which groups together spectrogram points coming from the same spatial location \cite{MandelEtAl2017}.  This information can be used to drive beamforming, which linearly combines multiple microphone channels into an estimate of the original signal that is optimal under some test-time criterion \cite{brandstein2013microphone}.  This optimality is typically based on properties of the signals or the spatial configuration of the recordings at test time, with no training ahead of time.

Another approach is to use a signal models trained using neural networks. Recent work on deep recurrent neural networks using the LSTM architecture \cite{hochreiter1997long} can achieve significant single-channel noise reduction \cite{6638947, weninger2015speech}, and so there is interest in using trainable deep-learning models to perform beamforming. This is especially useful for optimizing beamformers directly for automatic speech recognition \cite{xiao16, SainathEtAl2015}, although such optimization must happen at training time on a large corpus of training data.  Such models have difficulty generalizing across microphone arrays, including differences in number of microphones and array geometries, such as occurs between the AMI corpus \cite{carletta2007unleashing,4430116} and the CHIME challenge \cite{barker2015third}.  

In contrast to deep learning-based beamforming, spatial clustering is an unsupervised method for performing source separation, so it easily adapts across microphone arrays \cite{5200357, sawada2011underdetermined, bagchi15}.  Such methods group spectrogram points based on similarities in spatial properties, but are typically not able to take advantage of signal models, such as models of speech or noise. 

Developed by Mandel et al. \cite{5200357}, Model-based EM Source Separation and Localization (MESSL) is a system that computes time-frequency spectrogram masks for source separation as a byproduct of estimating the spatial location of the sources. It does so using the expectation maximization (EM) algorithm, iteratively refining the estimates of the spatial parameters of the audio sources and the spectrogram regions dominated by each source.

While MESSL utilizes spatial information to separate multiple sources, it does not model the content of the original signals.  This is an advantage when separating unknown sources, but performance can be improved when a model of the target source is available. The goal of this paper is to augment the capabilities of MESSL by adding a speech signal model based on neural networks trained to enhance the masks produced by MESSL.

In this paper we describe a novel method of combining single-channel LSTM-RNN-based speech enhancement into MESSL. We train a distinct LSTM model that uses the single-channel noisy audio to enhance the masks produced my MESSL.
To show how these methods enhance the speech of the CHiME-3 outdoor 6-channel audio, compared to baselines, we report the enhancement performance measured by PESQ score \cite{rix2001perceptual}, the SDR, SIR, and SAR scores from BSS Eval toolkit \cite{vincent2006performance}, as well as the WER as reported by the Kaldi toolkit \cite{Povey_ASRU2011} trained on a separate corpus, the indoor 8-channel AMI corpus.

\begin{table*}
\caption{Training Targets and their Associated Loss Function}
\centering
\label{tab:targets}
\begin{tabular}{lr@{$\,=\,$}lc}
\toprule
 & \multicolumn{2}{c}{Training Targets} &  Loss Functions\\ 
 \midrule
Ideal Amplitude (IA) Masks	&  $m_{ia}(\omega,t)$ &  $|s(\omega,t)|/|y(\omega,t)|$ & Binary Cross Entropy\\ 
Phase Sensitive (PA) Masks	&  $m_{ps}(\omega,t)$ & $\cos (\theta_{\omega, t}) \frac{|s(\omega,t)|}{|y(\omega,t)|}$ & Binary Cross Entropy\\ 
Magnitude Spectrum (MS) Approximation &  $m_{ma}(\omega,t)$ & $|s(\omega,t)|$ &  Mean-Squared Error \\ 
Phase-sensitive Spectrum (PS) Approximation & $m_{pa}(\omega,t)$ & $\cos(\theta_{\omega, t})|s(\omega,t)|$ &  Mean-Squared Error \\ 
\bottomrule
\end{tabular}
\end{table*}

\begin{figure*}[ht!]
    \centering
    \includegraphics[width=0.95\textwidth]{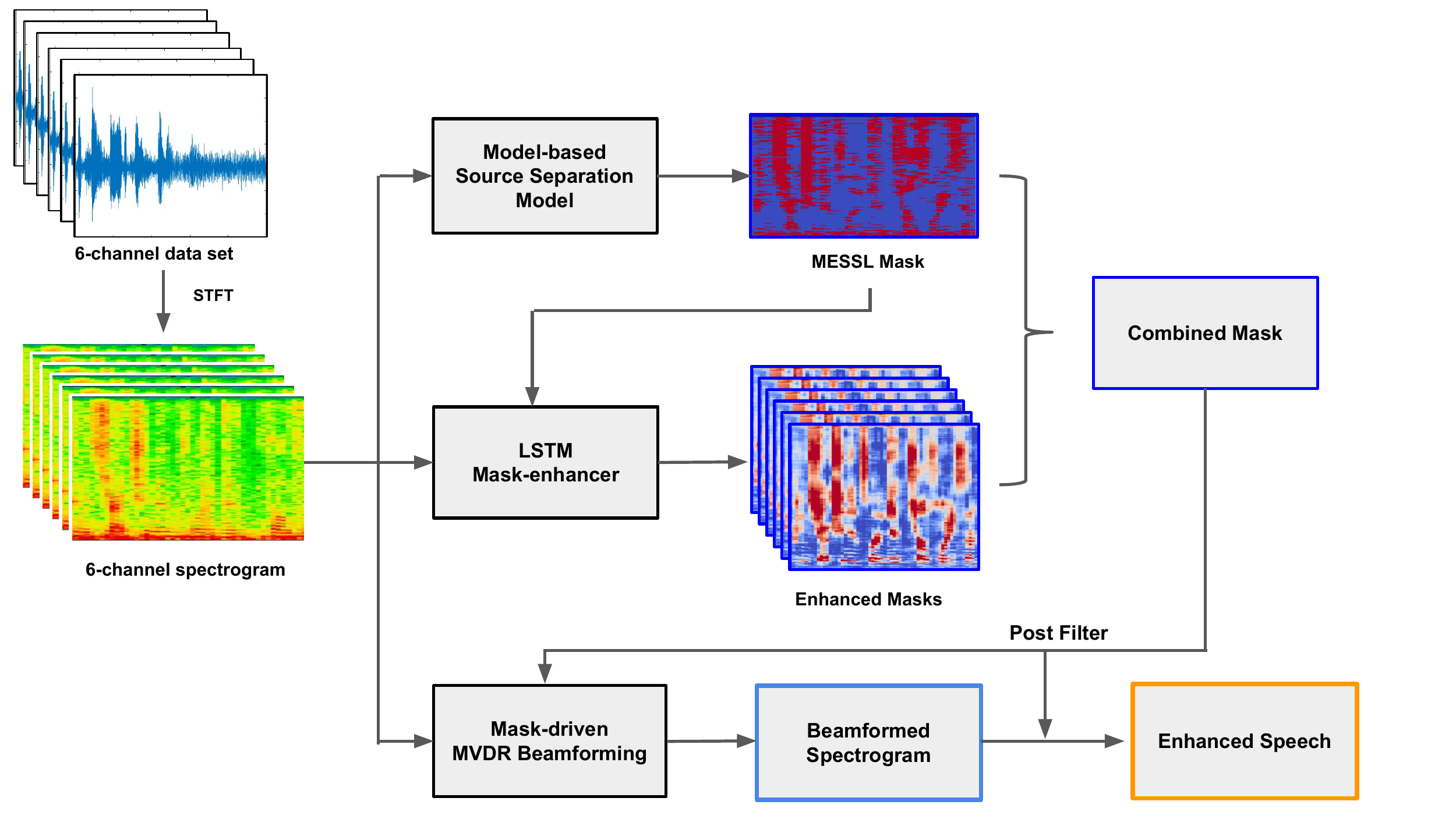}
    \caption{Multi-channel Spatial Clustering Based Time-Frequency Mask Enhancement System}
    \label{fig:framework}
\end{figure*}

\section{Related Work}
\label{sec:related}
Recently, Nugraha et al. \cite{nugraha2016multichannel} also studied multi-channel source separation using deep feedforward neural networks, using a multi-channel Gaussian model to combine the source spectrograms, to take advantage of the spatial information present in the microphone array. They explore the efficacy of different loss functions and other model hyper-parameters. One of their findings is that the standard mean-square error loss function performed close to the best. In contrast to our work they do not use spatial information that beamforming can give.

Pfeifenberger et al. \cite{pfeifenberger2017} proposed an optimal multi-channel filter which relies solely on speech presence probability. This speech-noise mask is predicted using a 2-layer feedforward neural network using features based on the leading eigenvector of the spatial covariance matrix of short time segments.  Using a single eigenvector makes the input to the DNN independent of the number of microphones, and thus adaptable to new microphone configurations.  It is trained on the simulated noisy data portion of CHIME-3. They show that this filter improves the PESQ score of the audio.  This approach uses an early fusion of the microphone channels before they are processed by the DNN, as opposed to our late fusion after the DNN processes each channel.

Heymann et al \cite{heymann2016neural,heymann2017} also study the combination of multi-channel beamforming with single-channel neural network model. Similar to ours, the proposed model consists of a bidirectional LSTM layer, followed by feedforward layers, in their case three. Of particular note is the companion paper by Boeddeker et al. \cite{boeddeker2017}, which derives the derivative of an eigenvalue problem involving complex-valued eigenvectors, allowing their system to propagate errors in the final SNR through the beamforming and back to the single-channel DNNs.  While we do not optimize our system in this end-to-end manner, the combination of MESSL with the per-channel DNNs may provide advantages in modeling the spatial information.

This paper builds upon previous work by the authors in \cite{mandel16b} and \cite{grezes17combining}, in which we propose two other methods of improving MESSL: a naive combination of the MESSL mask with the masks produced by a LSTM network trained to enhance noisy spectrograms, and using the LSTM-based masks to initialize the EM algorithm of MESSL. This previous work also describes a novel supervised-mvdr beamforming technique to obtain cleaner references for the CHiME-3 dataset.

\section{Methods}
\label{sec:methods}

\subsection{Training the Networks to Enhance the MESSL Masks}
To improve the quality of the binary masks produced by MESSL, we trained LSTM neural networks to enhance a MESSL mask when passing this mask and its associated noisy spectrogram through the network.
We tested four different training targets: ideal amplitude (IA) masks, phase sensitive (PA) masks, magnitude spectrum (MS) approximations and phase-sensitive (PS)	 spectrum approximations , based on work by Erdogan et al. \cite{erdogan2015phase}, as shown in Table~\ref{tab:targets}.
 

The LSTM operates on single-channel recordings.  Each channel in the multi-channel recording is processed independently and in parallel by the LSTM, following \cite{erdogan2016improved}.  In the single-channel setting, the short-time Fourier transform of the recorded noisy signal, $y(\omega,t)$ is assumed to be
\vspace{-1mm}
\begin{align}
y(\omega,t) = s(\omega,t) + n(\omega,t)
\end{align}
\vspace{-1mm}
where $s(\omega,t)$ is the (possibly reverberant) target speech and $n(\omega,t)$ is non-stationary additive noise. For the purposes of defining the targets and cost functions in Table~\ref{tab:targets}, let
\vspace{-1mm}
\begin{align}
\theta_{\omega, t} = \angle s(\omega,t) - \angle y(\omega,t)
\end{align}
\vspace{-1mm}
i.e., the phase difference between the target clean spectrogram and the input noisy spectrogram. In each case the network was configured to output a  $[0,1]$ valued mask $\hat{m}(\omega,t)$ for each frame of the input noisy spectrogram. 

For the masks targets, the network was trained to minimize the binary cross-entropy loss, while for the spectrum approximations targets the network was trained to minimize the mean-squared error. While in theory phase-sensitive masks may have negative values, causing problems with the cross-entropy loss function, in practice these were rare enough that we simply clipped those values to be 0.

For each training target type, we explored various hyper-parameter combinations: single or double bi-directional LSTM layers of size 256, 512, 1024 or 2048; merging of the bi-directional forward and backward outputs by summing, multiplying, averaging or concatenating; using a sigmoid or hard sigmoid (a piece-wise linear approximation of sigmoid that is faster to compute) for the output layer activation function. The exploration was done by randomly generating a network from the above 64 combinations and training it until the loss on the development set no-longer improved. For each training type, we report our best configuration in Table~\ref{tab:config}.

The spectrogram inputs were converted from a linear to decibel scale, and normalized to mean 0 and variance 1 at each frequency bin. The MESSL binary masks were passed through the logit function. To perform the computation and training of our LSTM neural networks, we used the KERAS python library \cite{chollet2015keras}, built upon the Tensorflow library \cite{tensorflow2015-whitepaper}.

\begin{figure}
    \centering
    \includegraphics[width=0.48\textwidth]{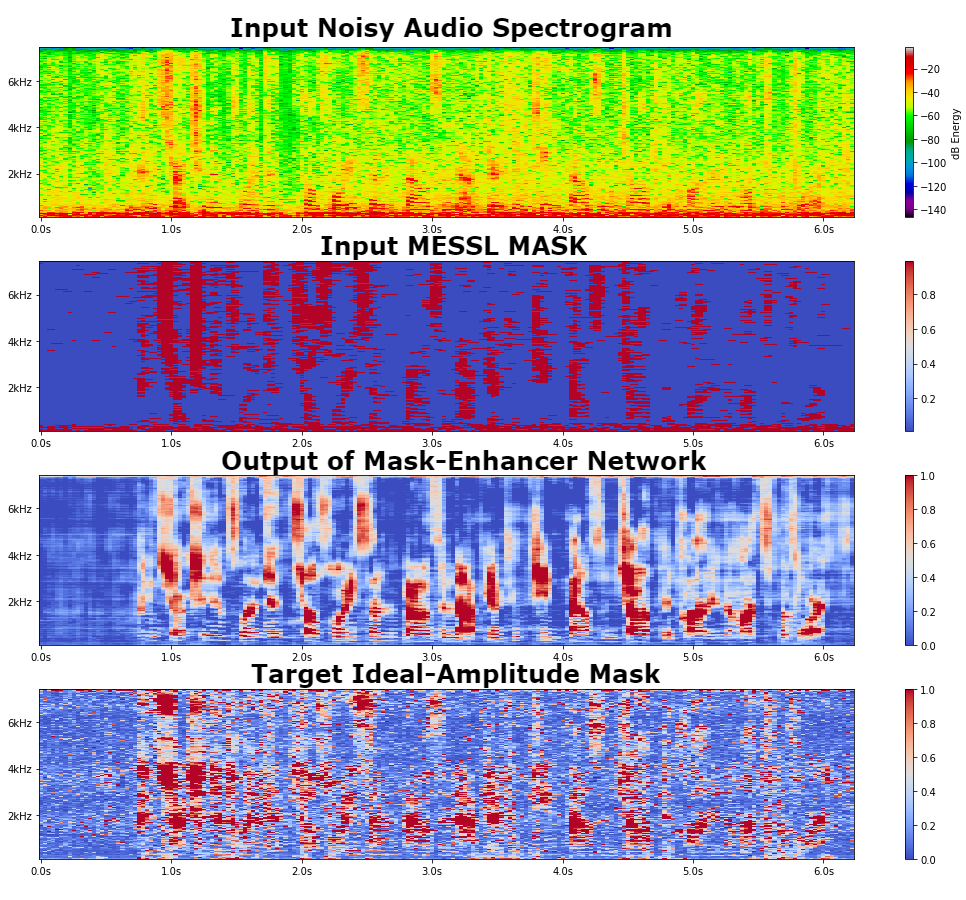}
    \caption{An example of the MA-trained Mask-Enhancer network using the noisy spectrogram from channel 1 in utterance F01\_050C0103\_BUS in the development set and its MESSL mask to produce an enhanced time-frequency mask.}
    \label{fig:me-ex}
\end{figure}

Figure~\ref{fig:me-ex} gives an example of how one of our networks has learned how to use the noisy spectrogram to refine a mask produced by MESSL.

\subsection{Enhancement of the MESSL Mask}

A flowchart illustrating the framework of our methods is shown in Figure~\ref{fig:framework}.
We extract six spectrograms from six-channel audio files using short-time Fourier transform (STFT). The window size is 1024 (64ms at 16kHz). We then use one of our models described above to enhance the mask produced by MESSL, using the six different channel spectrograms. Those six enhanced masks are combined into one by taking the maximum. We tried different ways of combining the MESSL mask and LSTM enhanced mask (average, maximum, minimum, or LSTM output only) into a final mask. A comparison of these combination methods is given in Table~\ref{tab:combo}
Then we use this final mask to estimate noise spatial covariances and perform mask-driven MVDR beamforming. We apply the same mask as a post filter onto the corresponding beamformed spectrogram and get the enhanced audio using the inversed short-time Fourier transform. This audio is then used to evaluate the quality of the model using the PESQ, SDR and WER metrics.

\section{Experiments and Results}
\label{sec:experiments}

\subsection{The CHiME-3 Corpus}
The CHiME-3 corpus features both live and simulated, 6-channel single speaker recordings from 12 different speakers (6 male, 6 female), in 4 different noisy environments: caf\'{e}, street junction, public transport and pedestrian area. 
In our work, we used the official data split, with 1600 real noisy utterances in the training set for training, 1640 real noisy utterances in the development set for validation. We did not use the simulated data to train our models. We tested our models on the proposed 2640 utterances in the test set, which contains audio both from real noisy recordings and simulated noisy recordings.
In order to perform speech recognition, we used the Kaldi toolkit trained on the AMI corpus, which features 8 microphones, recording overlapping speech in meeting rooms. These differences provide an additional challenge, but are essential to evaluating the generalization abilities of our model.

\subsection{Supervised MVDR Speech Reference}
Because the real subset of the CHiME-3 recordings were spoken in a noisy environment, it is not possible to provide a true clean reference signal for them.  Instead, an additional microphone was placed close to the talker's mouth to serve as a reference.  While this reference has a higher signal-to-noise ratio than the main microphones, it is not noise free.  In addition, because it is mounted close to the mouth, it contains sounds that are not desired in a clean output and actually could hurt ASR performance, namely pops, lip smacks, and other mouth noises.
In order to obtain a cleaner reference signal, we use the close mic signal as a frequency-dependent voice activity detector to control the MVDR beamforming of the signals from the array microphones, as described in \cite{grezes17combining}.

\subsection{Evaluation Metrics}
We evaluate the performance of our enhancement system in terms of both speech quality and intelligibility to a speech recognizer. For quality, we use the Signal-to-Distortion Ratio (SDR) from the BSS\_Eval toolkit \cite{vincent2006performance} and the Perceptual Evaluation of Speech Quality (PESQ) score \cite{Loizou2007}.  PESQ is measured in units of mean opinion score (MOS) between 0 and 5, higher being better.  For SDR, we used the source-based (not spatial image-based) scoring. For the simulated data, the reference signals were given by the booth recordings of CHiME-3. For the real data, the reference signals were given by the supervised MVDR for the target speech, and the approximation of the noise signals of the individual microphone channels were computed by subtracting the reference. Since we have no real ground truth audio for the real dataset, the SDR scores reported on that dataset should be taken with a grain of salt.  SDR is measured in decibels, with higher values being better.
PESQ is fairly accurate at predicting subjective quality scores for speech enhancement, but has the advantage for CHiME-3 of not requiring a reference for the noise sources. The supervised MVDR signal served as the speech reference for PESQ. 

We also evaluate the enhanced speech by Word Error Rate (WER) using Kaldi automatic speech recognizer. We train our Kaldi recognizer on AMI corpus. The training and test sets differ significantly in the number of microphones, array geometry, amount of reverberation, microphone array distance, amount and type of noise, speaking style, and vocabulary\cite{mandel2016multichannel}. After the training and testing setup, we can evaluate the performance of our enhancement system in reducing the mismatch between training and testing data. WER is measured in percent, with lower values being better.

\subsection{Baseline system} 

As a baseline, we use our own implementation of the method of Souden et al.~\cite{souden2011integrated}.  This approach generalizes improved minima-controlled recursive averaging \cite{cohen2002noise} to multichannel signals to estimate the speech presence probability.  This speech presence probability is then used to estimate the spatial covariance matrix of the noise, which is used to compute an MVDR beamformer.





\subsection{Results}

As detailed in section 3.1, we explored various architectures for each training target. We report the best architecture configurations in Table~\ref{tab:targets}, as measured by the loss on the CHiME-3 dev set.

\begin{table}
\caption{Best hyper-parameter settings found for each training target.  Multiple layer sizes indicates multiple layers.}
\label{tab:config}
\centering
\begin{tabular}{llll}
\toprule
\begin{tabular}[c]{@{}c@{}} Training \\ Target Type \end{tabular} &
\begin{tabular}[c]{@{}c@{}} Size of \\ LSTM Layer(s) \end{tabular} &
\begin{tabular}[c]{@{}c@{}} Bi-direction \\ Merge Mode \end{tabular} &
\begin{tabular}[c]{@{}c@{}} Output \\ Activation \end{tabular} \\ 
\midrule
IA & 512 & average & hard sigmoid \\ 
PS & (512,1024) & concatenation & sigmoid \\ 
MA & 512 & average & hard sigmoid \\ 
PA & (512, 2048) & concatenation & hard sigmoid  \\ 
\bottomrule
\end{tabular}
\end{table}

As detailed in section 3.2, we then tried various methods of combining the enhanced masks with the MESSL masks, from the best networks for each training target. As shown in Table~\ref{tab:combo}, we found that averaging the enhanced masks given by the model trained on ideal-amplitude targets produced the best results on the dev set.

\begin{table}
\caption{Comparison of WER for different methods of combining the enhanced mask with the MESSL mask, using the best performing model for each training target. (see Table~\ref{tab:config}).}
\label{tab:combo}
\centering
\begin{tabular}{llcccc}
\toprule
 &  & Avg & Min & Max & LSTM \\ 
 \midrule
\multicolumn{1}{l}{\multirow{2}{*}{IA}} & Dev & \textbf{19.3} & 19.7 & 21.3 & 20.1 \\ 
\multicolumn{1}{l}{} & Test & \textbf{32.6} & 32.3 & 33.1 & 32.0 \\ 
\multicolumn{1}{l}{\multirow{2}{*}{PS}} & Dev & 24.4 & 19.7 & 19.5 & 51.0 \\ 
\multicolumn{1}{l}{} & Test & 39.9 & 32.3 & 32.1 & 84.8 \\
\multicolumn{1}{l}{\multirow{2}{*}{MA}} & Dev & 29.1 & 19.7 & 20.0 & 69.6 \\ 
\multicolumn{1}{l}{} & Test & 45.3 & 32.3 & 32.4 & 85.7 \\ 
\multicolumn{1}{l}{\multirow{2}{*}{PA}} & Dev & 20.1 & 19.7 & 20.0 & 33.7 \\ 
\multicolumn{1}{l}{} & Test & 34.1 & 32.3 & 32.1 & 64.8 \\ 
\bottomrule
\end{tabular}
\end{table}


Finally, we fully evaluated our best model (IA, Avg) using the PESQ, SDR and WER metrics. The comparison to the baselines is shown in Tables~\ref{tab:res1} and \ref{tab:res2}. Compared to the method of Souden et al. \cite{souden2011integrated}, our mask-enhancer method achieves better scores across all three metrics and on both the dev and test dataset, over both the real and simulated data set, with one exception for the SDR score over the simulated test set. Compared to MESSL, our method improves the PESQ scores over the dev and test dataset, while achieving similar WER scores.

\begin{table}[ht!]
\caption{Results of PESQ and SDR comparing the best performing system from Table~\ref{tab:combo} with several baselines, over the simulated portion of the CHiME-3 eval and test data.}
\label{tab:res1}
\centering
\begin{tabular}{lllll}
\toprule
 & \multicolumn{2}{c}{PESQ (MOS)} & \multicolumn{2}{c}{SDR (dB)}  \\ 
 (SIMU data)& Dev & Test & Dev & Test  \\ 
 \midrule
\multicolumn{1}{l}{MESSL \cite{5200357}} & 3.18 &  3.10 & 6.00 & 2.38\\
\multicolumn{1}{l}{Souden \cite{souden2011integrated}} & 2.31 & 2.44 & 3.92 & 4.35  \\
\multicolumn{1}{l}{\textbf{Mask Enhancer}} & 3.15  & 3.13 & 6.24 & 2.57  \\ 
\bottomrule
\end{tabular}
\end{table}

\begin{table}[ht!]
\caption{Results of all metrics comparing the best performing system from Table~\ref{tab:combo} with several baselines, over the real portion of the CHiME-3 eval and test data.}
\label{tab:res2}
\centering
\begin{tabular}{lllllll}
\toprule
 & \multicolumn{2}{c}{PESQ (MOS)} & \multicolumn{2}{c}{SDR (dB)} & \multicolumn{2}{c}{WER (\%)} \\ 
 (REAL data)& Dev & Test & Dev & Test & Dev & Test \\ 
 \midrule
\multicolumn{1}{l}{MESSL \cite{5200357}} & 2.65 & 2.37 & 6.42 & 5.38 & 19.7 & 32.3 \\
\multicolumn{1}{l}{Souden \cite{souden2011integrated}} & 2.14 & 2.05 & 3.21 & 2.37  & 37.4 & 52.3 \\
\multicolumn{1}{l}{\textbf{Mask Enhancer}} &2.73  & 2.47 & 7.07 & 5.97 & 19.3 & 32.6 \\ 
\bottomrule
\end{tabular}
\end{table}

\section{Conclusion and Future Work}
\label{sec:conclusion}

In this paper we propose a novel method to adapt parallel single-channel LSTM-based enhancement to multi-channel audio, combining the speech-signal modeling power of the LSTM neural network with the spatial clustering power of MESSL, further enhancing the audio. We show that this method can help MESLL improve the quality of audio, with similar intelligibility.

Our future work will continue to explore different ways of integrating the LSTM speech-signal model with MESSL. Preliminary results show that the spatial information is more valuable than the single-channel speech information with respect to WER. To further test this hypothesis,the next step is to integrate a mask cleaning LSTM model in each loop of MESSL's EM algorithm, i.e use the mask enhancer model to clean the MESSL masks before the estimation of the spatial parameters.


\section{Acknowledgements}

This material is based upon work supported by the Alfred P Sloan foundation and the National Science Foundation (NSF) under Grant No. IIS-1409431. Any opinions, findings, and conclusions or recommendations expressed in this material are those of the author(s) and do not necessarily reflect the views of the NSF.

\newpage

{\footnotesize
\bibliographystyle{IEEEbib}
\bibliography{refs}

\begin{thebibliography}{10}

\bibitem{achieving-human-parity-conversational-speech-recognition-2}
Wayne Xiong, Jasha Droppo, Xuedong Huang, Frank Seide, Mike Seltzer, Andreas
  Stolcke, Dong Yu, and Geoffrey Zweig,
\newblock ``Achieving human parity in conversational speech recognition,''
\newblock Tech. {R}ep., February 2017.

\bibitem{MandelEtAl2017}
Michael~I Mandel, Shoko Araki, and Tomohiro Nakatani,
\newblock ``Multichannel clustering and classification approaches,''
\newblock in {\em Audio Source Separation and Speech Enhancement}, Emmanuel
  Vincent, Tuomas Virtanen, and Sharon Gannot, Eds., chapter~12. Wiley, 2017,
\newblock To appear.

\bibitem{brandstein2013microphone}
Michael Brandstein and Darren Ward,
\newblock {\em Microphone arrays: signal processing techniques and
  applications},
\newblock Springer Science \& Business Media, 2013.

\bibitem{hochreiter1997long}
Sepp Hochreiter and J{\"u}rgen Schmidhuber,
\newblock ``Long short-term memory,''
\newblock {\em Neural computation}, vol. 9, no. 8, pp. 1735--1780, 1997.

\bibitem{6638947}
A.~Graves, A.~r.~Mohamed, and G.~Hinton,
\newblock ``Speech recognition with deep recurrent neural networks,''
\newblock in {\em 2013 IEEE International Conference on Acoustics, Speech and
  Signal Processing}, May 2013, pp. 6645--6649.

\bibitem{weninger2015speech}
Felix Weninger, Hakan Erdogan, Shinji Watanabe, Emmanuel Vincent, Jonathan
  Le~Roux, John~R Hershey, and Bj{\"o}rn Schuller,
\newblock ``Speech enhancement with lstm recurrent neural networks and its
  application to noise-robust asr,''
\newblock in {\em International Conference on Latent Variable Analysis and
  Signal Separation}. Springer, 2015, pp. 91--99.

\bibitem{xiao16}
Xiong Xiao, Shinji Watanabe, Hakan Erdogan, Liang Lu, John Hershey, Michael~L
  Seltzer, Guoguo Chen, Yu~Zhang, Michael Mandel, and Dong Yu,
\newblock ``Deep beamforming networks for multi-channel speech recognition,''
\newblock in {\em Proceedings of the {IEEE} International Conference on
  Acoustics, Speech, and Signal Processing}. mar 2016, pp. 5745--5749, IEEE.

\bibitem{SainathEtAl2015}
T.~N. Sainath, R.~J. Weiss, A.~Senior, K.~W. Wilson, and O.~Vinyals,
\newblock ``Learning the speech front-end with raw waveform cldnns,''
\newblock 2015.

\bibitem{carletta2007unleashing}
Jean Carletta,
\newblock ``Unleashing the killer corpus: experiences in creating the
  multi-everything ami meeting corpus,''
\newblock {\em Language Resources and Evaluation}, vol. 41, no. 2, pp.
  181--190, 2007.

\bibitem{4430116}
S.~Renals, T.~Hain, and H.~Bourlard,
\newblock ``Recognition and understanding of meetings the ami and amida
  projects,''
\newblock in {\em 2007 IEEE Workshop on Automatic Speech Recognition
  Understanding (ASRU)}, Dec 2007, pp. 238--247.

\bibitem{barker2015third}
Jon Barker, Ricard Marxer, Emmanuel Vincent, and Shinji Watanabe,
\newblock ``The third chime speech separation and recognition challenge:
  Dataset, task and baselines,''
\newblock in {\em Automatic Speech Recognition and Understanding (ASRU), 2015
  IEEE Workshop on}. IEEE, 2015, pp. 504--511.

\bibitem{5200357}
M.~I. Mandel, R.~J. Weiss, and D.~P.~W. Ellis,
\newblock ``Model-based expectation-maximization source separation and
  localization,''
\newblock {\em IEEE Transactions on Audio, Speech, and Language Processing},
  vol. 18, no. 2, pp. 382--394, Feb 2010.

\bibitem{sawada2011underdetermined}
Hiroshi Sawada, Shoko Araki, and Shoji Makino,
\newblock ``Underdetermined convolutive blind source separation via frequency
  bin-wise clustering and permutation alignment,''
\newblock {\em IEEE Transactions on Audio, Speech, and Language Processing},
  vol. 19, no. 3, pp. 516--527, 2011.

\bibitem{bagchi15}
Deblin Bagchi, Michael~I Mandel, Zhongqiu Wang, Yanzhang He, Andrew Plummer,
  and Eric Fosler-Lussier,
\newblock ``Combining spectral feature mapping and multi-channel model-based
  source separation for noise-robust automatic speech recognition,''
\newblock in {\em Proceedings of the {IEEE} Workshop on Automatic Speech
  Recognition and Understanding}, 2015.

\bibitem{rix2001perceptual}
Antony~W Rix, John~G Beerends, Michael~P Hollier, and Andries~P Hekstra,
\newblock ``Perceptual evaluation of speech quality (pesq)-a new method for
  speech quality assessment of telephone networks and codecs,''
\newblock in {\em Acoustics, Speech, and Signal Processing, 2001.
  Proceedings.(ICASSP'01). 2001 IEEE International Conference on}. IEEE, 2001,
  vol.~2, pp. 749--752.

\bibitem{vincent2006performance}
Emmanuel Vincent, R{\'e}mi Gribonval, and C{\'e}dric F{\'e}votte,
\newblock ``Performance measurement in blind audio source separation,''
\newblock {\em IEEE transactions on audio, speech, and language processing},
  vol. 14, no. 4, pp. 1462--1469, 2006.

\bibitem{Povey_ASRU2011}
Daniel Povey, Arnab Ghoshal, Gilles Boulianne, Lukas Burget, Ondrej Glembek,
  Nagendra Goel, Mirko Hannemann, Petr Motlicek, Yanmin Qian, Petr Schwarz, Jan
  Silovsky, Georg Stemmer, and Karel Vesely,
\newblock ``The kaldi speech recognition toolkit,''
\newblock in {\em IEEE 2011 Workshop on Automatic Speech Recognition and
  Understanding}. Dec. 2011, IEEE Signal Processing Society,
\newblock IEEE Catalog No.: CFP11SRW-USB.

\bibitem{nugraha2016multichannel}
Aditya~Arie Nugraha, Antoine Liutkus, and Emmanuel Vincent,
\newblock ``Multichannel audio source separation with deep neural networks,''
\newblock {\em IEEE/ACM Transactions on Audio, Speech, and Language
  Processing}, vol. 24, no. 9, pp. 1652--1664, 2016.

\bibitem{pfeifenberger2017}
Lukas Pfeifenberger, Matthias Zohrer, and Franz Pernkopf,
\newblock ``Dnn-based speech mask estimation for eigenvector beamforming,''
\newblock in {\em Acoustics, Speech and Signal Processing (ICASSP), 2017 IEEE
  International Conference on}. 2017, IEEE SigPort.

\bibitem{heymann2016neural}
Jahn Heymann, Lukas Drude, and Reinhold Haeb-Umbach,
\newblock ``Neural network based spectral mask estimation for acoustic
  beamforming,''
\newblock in {\em Acoustics, Speech and Signal Processing (ICASSP), 2016 IEEE
  International Conference on}. IEEE, 2016, pp. 196--200.

\bibitem{heymann2017}
Jahn Heymann, Lukas Drude, Christoph Boeddeker, Patrick Hanebrink, and Reinhold
  Haeb-Umbach,
\newblock ``Beamnet: End-to-end training of a beamformer-supported
  multi-channel asr system,''
\newblock in {\em Acoustics, Speech and Signal Processing (ICASSP), 2017 IEEE
  International Conference on}, 2017.

\bibitem{boeddeker2017}
Christoph Boeddeker, Patrick Hanebrink, Jahn Heymann, Drude Lukas, and Reinhold
  Haeb-Umbach,
\newblock ``Optimizing neural-network supported acoustic beamforming by
  algorithmic differentiation,''
\newblock in {\em Acoustics, Speech and Signal Processing (ICASSP), 2017 IEEE
  International Conference on}, 2017.

\bibitem{mandel16b}
Michael~I Mandel and Jon~P Barker,
\newblock ``Multichannel spatial clustering for robust far-field automatic
  speech recognition in mismatched conditions,''
\newblock in {\em Proceedings of Interspeech}, 2016, pp. 1991--1995.

\bibitem{grezes17combining}
Felix Grezes, Ni~Zhaoheng, Trinh Viet~Anh, and Michael Mandel,
\newblock ``Combining spatial clustering with lstm speech models for
  multichannel speech enhancement,''
\newblock in {\em Interspeech}, 2017,
\newblock Submitted.

\bibitem{erdogan2015phase}
Hakan Erdogan, John~R Hershey, Shinji Watanabe, and Jonathan Le~Roux,
\newblock ``Phase-sensitive and recognition-boosted speech separation using
  deep recurrent neural networks,''
\newblock in {\em Acoustics, Speech and Signal Processing (ICASSP), 2015 IEEE
  International Conference on}. IEEE, 2015, pp. 708--712.

\bibitem{erdogan2016improved}
Hakan Erdogan, John~R Hershey, Shinji Watanabe, Michael Mandel, and Jonathan
  Le~Roux,
\newblock ``Improved mvdr beamforming using single-channel mask prediction
  networks,''
\newblock in {\em Proc. INTERSPEECH}, 2016.

\bibitem{chollet2015keras}
Fran\c{c}ois Chollet,
\newblock ``Keras,'' \url{https://github.com/fchollet/keras}, 2015.

\bibitem{tensorflow2015-whitepaper}
Mart\'{\i}n Abadi and et~al,
\newblock ``{TensorFlow}: Large-scale machine learning on heterogeneous
  systems,'' 2015,
\newblock Software available from tensorflow.org.

\bibitem{Loizou2007}
Philipos~C Loizou,
\newblock {\em Speech Enhancement: Theory and Practice (Signal Processing and
  Communications)},
\newblock CRC, 2007.

\bibitem{mandel2016multichannel}
Michael~I Mandel and Jon~P Barker,
\newblock ``Multichannel spatial clustering for robust far-field automatic
  speech recognition in mismatched conditions,''
\newblock {\em Interspeech 2016}, pp. 1991--1995, 2016.

\bibitem{souden2011integrated}
Mehrez Souden, Jingdong Chen, Jacob Benesty, and Sofiene Affes,
\newblock ``An integrated solution for online multichannel noise tracking and
  reduction,''
\newblock {\em IEEE Transactions on Audio, Speech, and Language Processing},
  vol. 19, no. 7, pp. 2159--2169, 2011.

\bibitem{cohen2002noise}
Israel Cohen and Baruch Berdugo,
\newblock ``Noise estimation by minima controlled recursive averaging for
  robust speech enhancement,''
\newblock {\em IEEE signal processing letters}, vol. 9, no. 1, pp. 12--15,
  2002.

\end{thebibliography}
}

\end{document}